\documentclass[twocolumn, amssymb, aps, pra]{revtex4-1}
\usepackage{natbib}
\usepackage{amsmath}
\usepackage{graphicx}
\usepackage{verbatim}
\usepackage{mathrsfs}
\usepackage{paralist}
\usepackage[normalem]{ulem}
\usepackage{color}

\begin{document}

\title[From the Jaynes-Cummings-Hubbard to the Dicke model]{From the Jaynes-Cummings-Hubbard to the Dicke model}

\author{S. Schmidt $^1$, G. Blatter $^1$, J. Keeling $^2$}
\address{$^1$ Institute for Theoretical Physics, ETH Zurich, CH-8093 Zurich, Switzerland}
\address{$^2$ Scottish Universities Physics Alliance, School of Physics and Astronomy, University of St Andrews,
St Andrews KY16 9SS, United Kingdom}
\email{Corresponding author\quad E-mail:\,\textsf{schmidts@phys.ethz.ch}}

\begin{abstract}
We discuss the Jaynes-Cummings-Hubbard model (JCHM) describing the superfluid-Mott insulator transition of polaritons (i.e., dressed
photon-qubit states) in coupled qubit-cavity arrays in the crossover from strong to weak correlations.
In the strongly correlated regime the phase diagram and the elementary excitations of lattice polaritons near the Mott lobes
are calculated analytically using a slave boson theory (SBT).
The opposite regime of weakly interacting polariton superfluids is described by a weak-coupling mean-field theory (MFT)
for a generalised multi-mode Dicke model.
We show that a remarkable relation between the two theories exists in the limit of large photon bandwidth and large negative detuning,
i.e., when the nature of polariton quasiparticles becomes qubit-like. In this regime, the weak coupling theory predicts the existence
of a single Mott lobe with a change of the universality class of the phase transition at the tip of the lobe, in perfect agreement with
the slave-boson theory. Moreover, the spectra of low energy excitations, i.e., the sound velocity of the Goldstone mode and the 
gap of the amplitude mode match exactly as calculated from both theories.
\end{abstract}
\maketitle

\section{Introduction}

The Jaynes-Cummings Hubbard model has been introduced by Greentree et al. \cite{Greentree2006}
to describe a possible superfluid-Mott insulator transition
of polaritons, i.e., quasiparticles of light and matter, in a coupled qubit-cavity array \cite{Greentree2006,Hartmann2006,Angelakis2007}. 
In this model a single photonic mode is strongly coupled 
to a two-level system (2LS) in each cavity and photons can hop between cavities. 
Hereby, the qubits introduce a nonlinearity into the system, 
which gives rise to effective repulsive photon-photon interactions.
The competition between repulsion (localization), and the photon hopping 
between cavities (delocalization) leads to an equilibrium 
quantum phase diagram featuring Mott lobes reminiscent of those of ultracold atoms 
in optical lattices as described by the Bose-Hubbard model ~\cite{Fisher1989}. 

In the strongly correlated regime of the JCHM, where the on-site repulsion dominates, 
photons become localised in a Mott-like state due to strong effective interactions. 
This extreme many-body state of light has been the subject of intense theoretical investigations (for recent reviews, see \cite{Houck2012,Schmidt2013}).
The quantum phase diagram and elementary excitations of the JCHM have been calculated 
using various methods, e.g., decoupling mean-field approximation \cite{Greentree2006}, DMRG ~\cite{Rossini2007,Rossini2008}, 
variational cluster approximation \cite{Aichhorn2008,Knap2010,Knap2011a}, strong coupling expansion \cite{Schmidt2009,Koch2009a,Schmidt2010,Nietner2012}
and Quantum Monte Carlo simulations \cite{Pippan2009,Hohenadler2011,Hohenadler2012}.
It was shown that the $U(1)$ symmetry breaking phase transition of the JCHM is in the same universality class
as the phase transition described by the BHM \cite{Schmidt2009,Koch2009a,Hohenadler2011}. Major differences in the shape of the Mott lobes and the
number of elementary excitations arise due to the $\sqrt{N}$-nonlinearity of the JCHM (as compared to a Kerr-like nonlinearity for the BHM)
and the composite nature of polariton quasiparticles (in the JCHM $N$ denotes the number of polaritons per site).
Possible experimental realizations of the JCHM include, e.g., cavity/circuit QED systems based on superconducting qubits in transmission line resonators \cite{Houck2012,Schmidt2013} and phonon-polaritons in trapped ion systems \cite{Ivanov2009,Hohenadler2012}.

In the regime of strong hopping and weak correlations, polaritons in a coupled qubit-cavity array
are expected to form a weakly interacting superfluid state. A BEC of weakly interacting polariton quasiparticles has already 
been observed experimentally with exciton-polaritons in a quantum well coupled to a semiconductor microcavity, formed by two Bragg mirrors~\cite{Kasprzak,Balili}. 
Spectacular experimental advances showed the existence of superfluidity \cite{Amo,Utsunomiya}, 
quantized vortices \cite{Lagoudakis}, and quantum solitons \cite{Amo2} in these systems. 
A comprehensive review of these experiments, and their relation to other
``quantum fluids of light'' can be found in the recent review~\cite{Carusotto2013a}. One approach to modelling polariton BEC's has been to use a
 generalised Dicke model in which many
two level systems, representing the presence or absence of an exciton
at a given site,
coherently couple to the multimode spectrum of a large cavity with quadratic photon dispersion \cite{Eastham00,Eastham2001,KE04,KE05}.
The transition from a normal to a superfluid state then corresponds to the equilibrium superradiance transition as originally predicted
for the single-mode Dicke model by Hepp and Lieb \cite{dicke54,Hepp:Super,wang73,hepp73:pra}.

Interestingly, the JCHM model is equivalent to the generalized Dicke model for small photon wave vectors. 
In this case the lattice dispersion can
be expanded yielding a quadratic photon dispersion with an effective mass that is inversely proportional 
to the hopping strength in the JCHM. 
In this paper we investigate in detail the connection between these two models. 
In particular, we show that a weak coupling mean-field theory for the generalised Dicke model
predicts the existence of Mott lobes and a change of the universality class of the phase transition at the tip of the lobe. 
This is in perfect agreement with results for the JCHM as obtained from a  slave-boson theory \cite{Schmidt2010}. 
In particular, the phase diagram and the elementary excitations at the phase boundary and inside the Mott lobes match exactly.
This result is in strong contrast to similar weak-coupling approaches for the Bose-Hubbard model, which 
fail to predict the existence of a superfluid-Mott insulator transition \cite{Oosten2001}.

Our paper is organised as follows:
In section \ref{sec:jchm} we introduce the Jaynes-Cummings Hubbard model and discuss its solution in the limit
of vanishing hopping. 
In section \ref{sec:slaveboson} we present the details of the slave-boson theory 
and discuss the phase diagram and the elementary excitations of the JCHM in the strongly correlated regime.
In section \ref{sec:mft} we map the JCHM to a generalised Dicke model and
show that a weak coupling mean-field theory also predicts the existence of a superfluid-Mott insulator transition. 
In section \ref{sec:connection}, we discuss the connection between the JCHM and the Dicke model at zero and finite temperatures. 
We conclude with a summary and outlook in section \ref{sec:conclusion}.

\section{The Jaynes-Cummings-Hubbard model (JCHM)}
\label{sec:jchm}

The Hamiltonian of the JCHM is given by
\begin{equation}
\label{jchm0}
H=\sum_i h^{\rm JC}_i - J \sum_{\langle i j \rangle} a^\dagger_i a_j - \mu N \,,
\vspace{-0.1cm}
\end{equation}
where $h^{\rm JC}_i$ denotes the local Jaynes-Cummings Hamiltonian 
\begin{eqnarray}
h^{\rm JC}_i = \omega_c\, a^\dagger_i a_i + \omega_x \sigma_{i}^+\sigma_{i}^-  +  g (\sigma_{i}^+ a_i +\sigma_{i}^- a^\dagger_i)
\end{eqnarray}
with site index $i$, boson creation (annihilation) operators $a_i^{(\dagger)}$ and qubit raising (lowering) operators $\sigma_i^{+(-)}$. 
The bosonic mode frequency is $\omega_c$, the two qubit levels are separated by the energy $\omega_x$ and the coupling is given by $g$ (we set $\hbar=1$).  

The on-site eigenstates of the Jaynes-Cummings Hamiltonian $h^{\rm JC}_i$ are labelled by the polariton number $n$ and upper/lower branch index $\sigma=\pm$. 
The mixed boson ($n,n-1$) - qubit ($g,e$) states define upper and lower polariton states
\begin{eqnarray}
\label{jcfunc}
|n +\rangle &=& \sin\theta_n |n\,,g\rangle + \cos\theta_n |(n-1)\,,e\rangle\,, \nonumber\\
|n -\rangle &=& \cos\theta_n |n\,,g\rangle -  \sin\theta_n |(n-1)\,,e\rangle\,,
\end{eqnarray}
where the weights are given by
\begin{eqnarray}
\sin\theta_n=\sqrt{\frac{1}{2}\left(1-\frac{\delta}{2\chi_n}\right)}
\end{eqnarray}
and
\begin{eqnarray}
\cos\theta_n=\sqrt{\frac{1}{2}\left(1+\frac{\delta}{2\chi_n}\right)}
\end{eqnarray}
with $\chi_n=\sqrt{g^2 n + \delta^2/4}$ and the detuning parameter $\delta=\omega_x-\omega_c$
The mixing angle $\theta_n$ can be written more compactly as $\theta_n=\tan^{-1}[2 g \sqrt{n}/(2\chi_n-\delta)]$.
The corresponding eigenvalues are
\begin{equation}
\label{jcen}
\epsilon_n^{\sigma}=-(\mu-\omega_c) n + \delta/2 +\sigma\,\chi_n\,,\quad \sigma=\pm\,.
\end{equation}
The zero polariton state $|0\rangle\equiv|0-\rangle=|0\,,g\rangle$ is a special case with $\epsilon_0\equiv\epsilon_0^-=0$.\\

Due to the mixing in (\ref{jcfunc}), polaritons can be interpreted as dressed photons.
This dressed photon inherits the anharmonicity of the matter component, i.e., the qubit,
leading to a nonlinearity in the spectrum of the combined light-matter system, which is given by
the energy difference
\begin{eqnarray}
\label{eq:Hubbard-U}
U=(\epsilon^\pm_{n=2} - \epsilon^\pm_{n=1}) - (\epsilon^\pm_{n=1}-\epsilon_{n=0})\,.
\end{eqnarray}
The parameter $U$ describes the energy cost of adding a second photon to the cavity versus adding the first.
It quantifies the effective on-site repulsive interaction among photons and is thus sometimes also called the particle-hole gap or effective Hubbard-U.
The spectral shift in Eq.~(\ref{eq:Hubbard-U}) is largest for zero qubit-cavity detuning ($\delta=0$), where $U=g(2-\sqrt{2})$. It becomes vanishingly small for large detunings, i.e., $U=\mathcal{O}(g^4/\delta^3)$. In this dispersive regime, photons and qubits barely interact with each other.

The second term in (\ref{jchm0}) describes the delocalization of bosons over the whole lattice due to hopping between nearest neighbour sites with amplitude $J$. It competes with an effective on-site repulsion as given by $U$, which is mediated by the coupling $g$.

In order to calculate the phase diagram of the JCHM it is convenient to work in the grand-canonical formalism. 
The global $U(1)$ symmetry of the JCHM preserves the total number of polaritons $N=\sum_i ( a^\dagger_i a_i + \sigma_{i}^+\sigma_{i}^-)$.
We thus introduce a chemical potential $\mu$ in the third term of (\ref{jchm0}), which fixes the number of polaritons.
While this might be justified for some experimental setups with large coherence times,
e.g., phonon-polaritons in trapped ion systems, it does not take into account the basic nature of most quantum optical applications: drive
and dissipation. Some recent works have started to address this issue and suggest that remnants of the equilibrium SF-MI transition remain
visible even in this strongly non-equilibrium situation \cite{Tomadin2010a,Liu2011,Nissen2012,Kuliatis2012,Grujic2012,Grujic2013}.
Note, that there have also been related studies of the driven, dissipative Bose-Hubbard model \cite{Hartmann2010, Boite2012}.
In this paper we assume that the system equilibrates and that the lifetime of the polariton quasiparticles is much larger than the measurement time.

\section{Strong correlations: Slave-Boson theory}
\label{sec:slaveboson}

An analytic strong-coupling theory for the phase diagram and the elementary excitations in the Mott phase of the JCHM
has been derived based on a linked-cluster expansion (LCA). This strong-coupling theory has been generalised to the superfluid 
phase using a slave-boson approach \cite{Schmidt2010}, which was previously applied to the BHM \cite{Huber2007}. 
Below we present the details of this formalism with explicit algebraic expressions
for the quantum phase diagram and the elementary excitations of the JCHM.

\subsection{Slave-boson formulation}
A convenient starting point for our slave-boson approach is the polariton representation ~\cite{Koch2009a} of the boson operator 
\begin{eqnarray}
a_i=\sum_{n \sigma \nu} f^{\sigma \nu}_n P^{\nu\dagger}_{i n-1} P^\sigma_{i n}
\end{eqnarray}
in terms of standard algebra operators $P^{\sigma\dagger}_{i n}=|n \sigma\rangle_{ii}\langle 0 |$ and matrix elements $f^{\sigma \nu}_n= \langle n-1\,\nu | a | n\, \sigma\rangle$
with $f^{\sigma \nu}_n=\left( \sqrt{n}+\sigma\,\nu\,\sqrt{n-1}\right)/2$ for $n>1$  ($f^{\sigma -}_1=1/\sqrt{2}$) at zero detuning ($\delta=0$). 
It is straightforward to show that the polariton operators obey bosonic commutation relations if the constraint (completeness relation)
\begin{equation}
\label{constraint}
\sum_{n\sigma} P^{\sigma\dagger}_{i n} P^\sigma_{in}=1
\end{equation}
is fullfilled at each site $i$. 
In this new basis the JCHM becomes
\begin{eqnarray}
\label{jchm}
H=&&\sum_i\sum_{n=0}^\infty\sum_\sigma \epsilon^\sigma_{n} {P^{\sigma\dagger}_{i n}} P^\sigma_{in}\\
&-& J\sum_{\langle ij\rangle}\sum_{n,n'=1}\mathop{\sum_{\scriptstyle\sigma,\sigma'}}_{\nu,\nu'}f^{\sigma\sigma'}_n f^{\nu\nu'}_{n'}\, P^{\sigma\dagger}_{i n} P^{\sigma'}_{i n-1} P^{\nu' \dagger}_{j n'-1} P^\nu_{j n'}\,.\nonumber
\end{eqnarray}
In ~\cite{Schmidt2009} we have shown that the presence of the upper polariton branch with $\sigma,\sigma' =+$ leads to additional high energy
conversion modes in the Mott phase with small spectral weight and bandwidth. We thus neglect the upper 
branch as well as particle conversion hopping (i.e, processes where a polariton hops to another site and at the same time changes its nature from
upper to lower or vice versa) from now on and drop the branch index $\sigma$. 
This leads to the simplified Hamiltonian
\begin{eqnarray}
\label{jchm2}
H=&&\sum_i\sum_{n=0}^\infty \epsilon_{n} P^{\dagger}_{i n} P_{in}\\
&-& J\sum_{\langle ij\rangle}\sum_{n,n'=1} f_n f_{n'}\, P^{\dagger}_{i n} P_{i n-1} P^{\dagger}_{j n'-1} P_{j n'}\nonumber\,.
\end{eqnarray}
with $\epsilon_{n}\equiv \epsilon^-_{n}$ and $f_{n}\equiv f^{--}_{n}$.

\subsection{Mean field theory}
\begin{figure}[t]
\centering
\includegraphics[width=0.45\textwidth,clip]{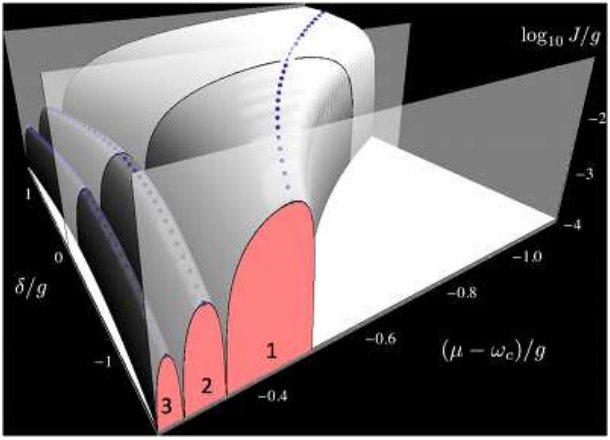}  
\caption{Quantum phase diagram for the JCHM as obtained from slave-boson theory, i.e., Eq.~(\ref{boundary}). Shown are the lowest three Mott lobes with polariton numbers $N=1,2,3$. Dotted lines represent the critical hopping strength's $J_c/g$ for which the chemical potential $\mu$ and detuning $\delta$ are chosen such as to fullfill particle-hole symmetry. Finite detuning $|\delta|>0$ decreases the critical hopping strength $J_c/g$ for $N>1$, but the lowest Mott lobe ($N=1$) steadily increases when tuning through the resonance ($\delta=0$). Figure taken with permission from ~\cite{Schmidt2010} (with minor modifications).}
\label{fig1}
\end{figure}

In order to calculate the phase boundary and static observables in the superfluid phase near
a Mott lobe with filling $n\geq 1$, we restrict the Hilbert space to states with $n$ and $n\pm1$ bosons
and make a Gutzwiller Ansatz for the ground-state wave function
\begin{equation}
|\psi\rangle=\hspace{-0.1cm}\prod_i\hspace{-0.1cm}\big[ \cos(\theta) P^\dagger_{i 0} \hspace{-0.05cm}+ \sin(\theta)( \sin(\chi) P^\dagger_{i -1} \hspace{-0.06cm}+\hspace{-0.05cm} \cos(\chi) P^\dagger_{i 1}) \big]|0\rangle\nonumber\\
\end{equation}
where we also dropped the index $n$ and changed the notation to $P^\dagger_{i\alpha}\equiv P^\dagger_{in+\alpha}$, $\epsilon_{n+\alpha}\equiv \epsilon_\alpha$, and $f_{n+\alpha}\equiv f_\alpha$. Note, that this variational wave function is normalized to unity and satisfies the completeness relation (\ref{constraint}) in the restricted Hilbertspace of $(n,n\pm1)$ lower polaritons.\\
The expectation value $\epsilon_{\rm var}=\langle \psi | H | \psi \rangle$ yields the variational energy
\begin{eqnarray}
\label{evar}
\epsilon_{\rm var}&=&\epsilon_0 \cos(\theta)^2+\sin(\theta)^2\left[ \epsilon_{-1}\sin(\chi)^2+\epsilon_{1}\cos(\chi)^2\right]\nonumber\\
&&- J D/2 \sin(2\theta)^2\left[ f_0 \cos(\chi) + f_{-1} \sin(\chi)\right]^2\,,
\end{eqnarray}
which has to be minimized with respect to the variational parameters $\theta$ and $\chi$. Here, $D$ denotes the dimension of a hypercubic lattice.
We obtain the relations
\begin{eqnarray}
\label{chi}
\tan(2\chi)=\frac{4 J D f_0 f_1 \cos(\theta)^2}{\epsilon_{-1}-\epsilon_1+2 J D (f_1^2-f_0^2) \cos(\theta)^2}
\end{eqnarray}
and
\begin{eqnarray}
\label{theta}
\cos(2\theta)=\frac{1}{2 J D} \frac{\epsilon_1 \cos(\chi)^2 +\epsilon_{-1}\sin(\chi)^2-\epsilon_0}{[f_0\sin(\chi)+f_1\cos(\chi)]^2}
\end{eqnarray}
The lobe boundaries are determined by the vanishing of the order parameter 
\begin{eqnarray}
\phi_c=\langle \psi | a | \psi\rangle=\sin(\theta)\left[f_0\sin(\chi)+f_{1}\cos(\chi)\right]^2/2.
\end{eqnarray}
Setting $\phi_c=0$ (i.e, $\theta=0$) in (\ref{chi}) and (\ref{theta}) and eliminating $\chi$ yields the relation
\begin{eqnarray}
\label{boundary}
\epsilon_{-1}-\epsilon_1=-Jz(f_1^2-f_0^2)\pm\sqrt{Q}
\end{eqnarray}
with
\begin{eqnarray}
Q=U^2-2Jz(f_0^2+f_1^2) U+J^2z^2(f_1^2-f_0^2)^2\,.
\end{eqnarray}
Eq.~(\ref{boundary}) constitutes an expression for the mean-field boundaries of the Mott lobes in the JCHM shown in Fig.~\ref{fig1}.
Here, we point out that the size of all Mott lobes with filling factor $n>1$ decrease for any finite detuning $|\delta|>0$, while the size of the lowest Mott lobe ($n=1$)
increases with negative detuning ($\delta<0$). Only in this latter case the nature of all polaritons in the lattice become qubit-like and are thus trivially localized.
Fig.~\ref{fig1} thus already suggests that a weak-coupling mean-field approach might be suitable for a description of the lowest Mott lobe in the limit of large hopping and 
negative detuning. We will further elaborate on this in section \ref{sec:connection}.
\subsection{Elementary excitations}

In order to find the elementary excitations we define a new set of operators ${\bf R}^\dagger=(G_i^\dagger, E_{1i}^\dagger, E_{2i}^\dagger)^T$, which is 
obtained from the original polariton basis ${\bf P}^\dagger=(P_{i 0}^\dagger, P_{i -1}^\dagger, P_{i 1}^\dagger)^T$
via a unitary transformation ${\bf R}^\dagger=T {\bf P}^\dagger$ with
\begin{eqnarray}
T=\left( \begin{array}{ccc} \cos(\theta) & \sin(\theta)\cos(\chi) & \sin(\theta)\sin(\chi) \\ 
-\sin(\theta) & \cos(\theta)\cos(\chi) & \cos(\theta)\sin(\chi) \\ 0 & -\sin(\chi) & \cos(\chi) \end{array} \right)\,.
\end{eqnarray}
The operator $G^\dagger$ creates a new vacuum state, i.e., the mean-field ground state $|\psi\rangle=\prod_i G_i^\dagger |0\rangle$,
and $E_{1i}^\dagger, E_{2i}^\dagger$ are orthogonal operators creating excitations above the ground-state.
We express the Hamiltonian in terms of these new operators and eliminate $G_i$ by using the constraint (\ref{constraint})
in the restricted Hilbert space 
\begin{eqnarray}
G_i\approx\sqrt{1-E_{1i}^\dagger E_{1i}-E_{2i}^\dagger E_{2i}}
\end{eqnarray}
Expanding the square root  everywhere in the Hamiltonian to quadratic order in $E^{(\dagger)}_{(1,2)i}$ yields,
after a Fourier transformation, an effective quadratic Hamiltonian
\begin{eqnarray}
H_{\rm eff}=\epsilon_{\rm var}+\sum_{\bf k} {\bf E}_{\bf k}^\dagger\, h_{{\rm eff},{\bf k}} \,{\bf E}_{\bf k}\,
\end{eqnarray}
where ${\bf E}=(E_{1{\bf k}},E_{2{\bf k}},E_{1-{\bf k}}^\dagger,E_{2-{\bf k}}^\dagger)^T$
and $h_{{\rm eff},{\bf k}}$ is a $4\times 4$ matrix
\begin{eqnarray}
\label{eq:sb-heff}
h_{{\rm eff},{\bf k}}=\left( \begin{array}{ccc} g & f \\ f & g \end{array} \right)\,,
\end{eqnarray}
with $f,g$ denoting $2\times 2$ matrices defined in the appendix.
The sum over ${\bf k}$ runs over the first Brioullin zone.
The effective Hamiltonian can be diagonalized by a bosonic Bogoliubov transformation yielding
\begin{equation}
H_{\rm eff}=\epsilon_{\rm var}+\epsilon_{\rm fluct}+\sum_{\alpha=\pm}\sum_{\bf k} \epsilon_\alpha({\bf k}) d_{\alpha{\bf k}}^\dagger d_{\alpha{\bf k}}
\end{equation}
with a fluctuation-generated correction of the ground-state energy
\begin{eqnarray}
\label{fluct}
\epsilon_{\rm fluct}=\mathcal{E}(\theta,\chi)+\sum_{\alpha=\pm}\sum_{\bf k} \epsilon_{\alpha}({\bf k})/2
\end{eqnarray}
and $d_{\alpha {\bf k}}^\dagger$ creating excitations with energy 
\begin{eqnarray}
\label{spectterms}
\epsilon_{\pm}({\bf k})=\sqrt{A({\bf k})\pm\sqrt{A({\bf k})^2- B({\bf k})}}\,
\end{eqnarray}
with expressions for $\mathcal{E}(\theta,\chi)$, $A({\bf k})$, and $B({\bf k})$ given in the appendix.

\begin{figure}[t]
\centering
\includegraphics[width=0.45\textwidth,clip]{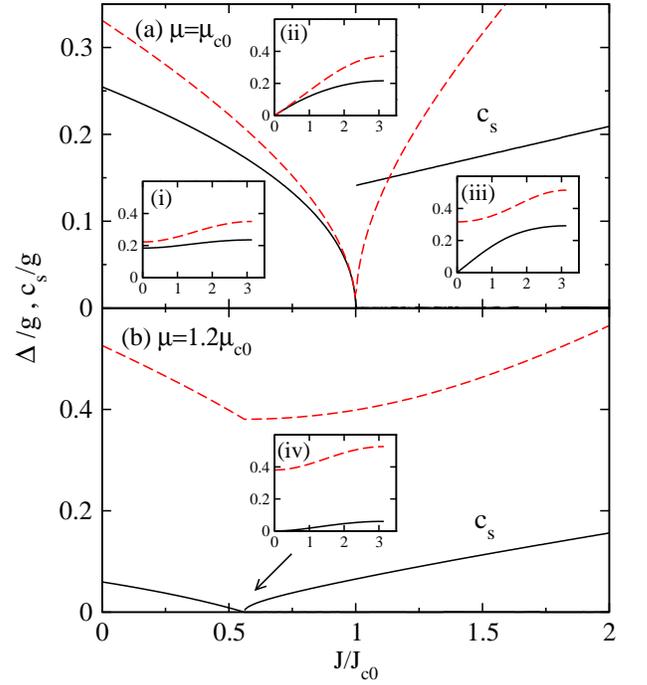}
\caption{\label{fig2} Elementary excitations of the JCHM as a function of the effective hopping strength $J/J_{c0}$ at zero detuning $\delta/g=0$ and for 
(a) $\mu=\mu_{c0}$ where $\mu_{c0}$ denotes the critical chemical potential at the tip of the lobe with critical hopping strength $J_c=J_{c0}$ (top figure) 
and (b) away from the tip at $\mu=1.2\mu_{c0}$ with  $J_c=0.566J_{c0}$ (bottom figure).\\
Shown are the gaps of particle (dashed) and hole (solid) modes in the Mott phase ($J<J_c$) and as well as the gaps 
of the Amplitude mode (dashed) and the sound velocity of the Goldstone mode (solid).
The insets show the corresponding excitation spectra at i) $J=0.5J_{c0}$ (in the Mott phase)
ii) $J=J_{c0}$ (at the tip of the lobe)  iii) $J=1.5J_{c0}$ (in the superfluid phase)  iv) $J=0.566J_{c0}$ (at the phase boundary away from the tip of the lobe).\\
At the phase boundary, the particle and hole mode of the Mott phase are identical with the Goldstone and Amplitude modes of the superfluid phase. 
At the tip of the lobe ii), where the polariton density can remain constant during the superfluid-insulator transition, the Amplitude mode becomes gapless and linear (its mass vanishes).
The sound velocity of the Goldstone mode remains non-zero, confirming a special point in the phase diagram with dynamical critical exponent $z=1$.
Away from the tip iv), the Amplitude mode remains gapped and the Goldstone mode becomes quadratic with a vanishing sound velocity corresponding
to a generic dynamical critical exponent $z=2$. Figure taken with permission from ~\cite{Schmidt2010} (with minor modifications).}
\end{figure}
In the Mott phase, the spectra can be written explicitely as 
\begin{eqnarray}
\label{spectrum}
\epsilon_{\pm}({\bf k})=\frac{1}{2}\left(\epsilon_{-1}-\epsilon_1+J\epsilon_{\bf k}(f_1^2-f_0^2)\pm\sqrt{Q({\bf k})}\right)
\end{eqnarray}
with
\begin{eqnarray}
Q({\bf k})=U^2-2UJ\epsilon_{\bf k}(f_1^2+f_0^2)+J^2\epsilon_{\bf k}^2(f_1^2-f_0^2)^2
\end{eqnarray}
and the single-particle spectrum
\begin{eqnarray}
\epsilon_{\bf k}=2\sum_{i=1}^D \cos(k_i)\,.
\end{eqnarray}
We thus obtain two gapped modes corresponding to particle/hole like excitations (see, Fig.~\ref{fig2}). 

In the superfluid phase, we obtain a gapless, linear Goldstone mode 
$\epsilon_-({\bf k}) = c_s |{\bf k}| + \mathcal{O}({\bf k}^2)$
with a finite sound velocity $c_s$.  At the phase boundary,
the sound velocity of this Goldstone mode vanishes except at the tip of the lobe,
where the sound velocity maintains a finite value different from zero.
This leads to a change of the dynamical critical exponent of the SF-MI transition from its generic value $z=2$ to $1$. The JCHM is thus in the same universality class as the BHM \cite{Schmidt2009,Koch2009a}. This has been confirmed
by large scale Quantum Monte-Carlo simulations \cite{Hohenadler2011}.

A second mode, the so-called amplitude or Higgs mode, generally remains gapped with
$\epsilon_+({\bf k})=\Delta_a + \mathcal{O}({\bf k}^2)$ except for the tip of the lobe, where the gap $\Delta_a$ vanishes. Thus, the amplitude mode at the tip of the lobe 
becomes linear consistent with a change of the dynamical critical exponent as discussed above.
For a detailed discussion of the excitation spectra we refer to the caption in Fig.~\ref{fig2}.

\section{Weak correlations: Bogoliubov-like theory}
\label{sec:mft}
The quantum phase transitions in the JCHM separates a phase with a
broken $U(1)$ symmetry from ``normal'' insulating states.  At finite
temperature, as for the Bose-Hubbard model, the insulating Mott lobes
join up to become a normal state (and the quantised occupation is
destroyed at finite temperature).  Viewed this way, there is a clear
relation between the phase transitions in the JCHM and the
Dicke-Hepp-Lieb~\cite{dicke54,Hepp:Super,wang73,hepp73:pra}
superradiance phase transition of the Tavis-Cummings
model~\cite{Tavis:TC}, which is frequently referred to as the Dicke
model (some authors make the distinction that the Dicke model
  contains also counter-rotating terms in the coupling between light
  and matter, however this naming convention is not followed by all
  authors).  This section discusses how theories developed for the
Dicke model can be used to understand the JCHM.  Surprisingly, this
reveals that the Dicke model shows Mott lobes, and that the
generalised Dicke model defined below contains a point, where the
universality class of the phase transition changes, just as for the
JCHM.

\subsection{Mapping to the Dicke model}

By Fourier transforming the photon operators to momentum space
\begin{eqnarray}
a_i=\frac{1}{\sqrt{N_s}}\sum_{\bf k} a_{\bf k} e^{i{\bf k}\cdot{\bf r}_i}
\end{eqnarray} 
the JCHM can be written as
\begin{eqnarray}
\label{dicke}
H=\sum_{\bf k} \tilde{\omega}_{\bf k} a_{\bf k}^\dagger a_{\bf k}+\sum_i \tilde{\omega_x}\sigma_i^+\sigma_i^- + \sum_{i{\bf k}}\frac{g_{i{\bf k}}}{\sqrt{N_s}}\left(\sigma_i^+ a_{\bf k}+{\rm h.c.}\right)\nonumber\\
\end{eqnarray}
where $\tilde{\omega}_{\bf k} = \tilde{\omega_c} - 2J\sum_{\alpha=1}^D \cos(k_\alpha)$ 
with $\tilde{\omega_c}=\omega_c-\mu$, $\tilde{\omega_x}=\omega_x-\mu$ and $g_{i {\bf k}} = g e^{i{\bf k}\cdot{\bf r}_i}$ ($N_s$ denotes the number of lattice sites).
This Hamiltonian represents a many-mode Dicke model, as studied in Refs.~\cite{KE04,KE05}.  The
case studied in those works, however, considered a quadratic photon
spectrum, equivalent to expanding the lattice dispersion for small
${\bf k}$ vectors yielding
\begin{eqnarray}
  \tilde{\omega}_{\bf k}=
  \tilde{\omega_c} - 2DJ + J {\bf k}^2
  \equiv
  J{\bf k}^2-\mu_D.
\end{eqnarray}
Here, we have defined a Dicke-model chemical potential $\mu_D = \mu +
2DJ - \omega_c$, such that $\mu_D<0$ is required for thermodynamic
stability.  It is similarly useful to define a Dicke-model detuning
$\delta_D =\delta + 2DJ$, measuring the detuning
between the 2LS energy and the bottom of the photon band so that
$\tilde{\omega}_x = \delta_D - \mu_D$.  With this quadratic expansion
the generalised Dicke model describes $N_s$
localised two-level systems coherently coupled to a continuum of
photonic modes with an effective photonic mass $1/2J$.

The quadratic expansion of the dispersion removes behaviour arising
when the bandwidth becomes small compared to other energy scales,
i.e., the Dicke model with quadratic dispersion corresponds to the JCHM
in the limit of large bandwidth $J$.  However, as discussed below,
even with this restriction, the first two Mott lobes can still be
reached. For the single mode Dicke model, in the $N_s\to\infty$ limit,
mean-field theory is exact, i.e., fluctuation corrections are
suppressed as $1/N_s$.  However, for the generalised Dicke model this is
not the case. Fluctuation corrections due to finite momentum photon
modes can shift the phase boundary~\cite{KE04,KE05} and change the
critical behaviour from mean-field to that of the XY model.  This
shift (and the size of the fluctuation dominated regime, as determined
by the Ginzburg criterion) is, however, small if the density of states
of finite momentum modes $\propto m^{D/2} = (2J)^{-D/2}$ is small.
As such, the limit in which the JCHM and Dicke models match is also
the limit in which fluctuation corrections to mean field theory become
negligible.  In the following, we thus first discuss the mean field
theory and the spectrum of fluctuations about this point. The
following section relates these ideas to the effect of fluctuations on
the phase boundary and further connections between the JCHM and Dicke
model phase diagrams.

\subsection{Mean-field theory of the Dicke model}

We now first consider a mean-field approximation for the single photon mode
with zero wave vector, i.e., $a_{\bf k=0}$. As first discussed
by Hepp and Lieb~\cite{Hepp:Super,hepp73:pra}, for a single mode,
there is a transition to a superradiant state (i.e., a superfluid state
of the JCHM with broken $U(1)$ symmetry) if
\begin{equation}
  \label{eq:1}
  g^2 >
  \frac{\tilde{\omega}_{\bf k=0} \tilde{\omega}_x}{%
    \tanh\left(\beta \tilde{\omega}_x/2\right)}
  =
  -\mu_D
  \frac{(\delta_D-\mu_D)}{\tanh\left(\frac{\beta}{2}(\delta_D-\mu_D)\right)}.
\end{equation}
Here, $\beta=1/T$ denotes the inverse temperature $T$ given in units of the Boltzman
constant $k_B$.
The original idea of the superradiant phase transition for the ground
state of the Dicke model was later questioned by Rzazewski {\it et
  al.}~\cite{Rzazewski1975} who pointed out that diamagnetic $A^2$
terms prevent the phase transition of two-level systems coupled to a
photon mode, leading to a ``no-go theorem'' for the superradiance
transition.  The presence of a chemical potential $\mu_D$ avoids this
no-go theorem: increasing the density of excitations by increasing
$\mu_D$ reduces the critical $g$ to a regime where diamagnetic $A^2$
terms can be neglected~\cite{Eastham00}.

This is particularly clear at $T=0$.  In this case the mean-field
self-consistency equation (gap equation) for $\psi_0=\langle a_{\bf k=0}\rangle$
becomes
\begin{eqnarray}
 \tilde{\omega}_{\bf k=0}|\psi_0|= \frac{g^2}{E}|\psi_0|\,,
\end{eqnarray}
where $E=\sqrt{\tilde{\omega}_x^2+4 g^2|\psi_0|^2}$.
Thus, the photonic condensate is given by
\begin{eqnarray}
  |\psi_0|^2=\frac{1}{4}\left(
    \frac{g^2}{\tilde{\omega}_{\bf k=0}^2}-\frac{\tilde{\omega}_x^2}{g^2}
  \right)
\end{eqnarray}
The transition from superfluid to normal phase is signalled by the
vanishing of the order parameter, i.e., $\psi_0=0$, corresponding to
$g^2 = \tilde{\omega}_{\bf k=0} |\tilde{\omega}_x| = - \mu_D |\delta_D -
\mu_D|$ (the modulus sign appearing here can be understood from the
$T \to 0$ limit of equation~(\ref{eq:1})).  

\begin{figure}[t]
  \centering
\includegraphics[width=0.45\textwidth,clip]{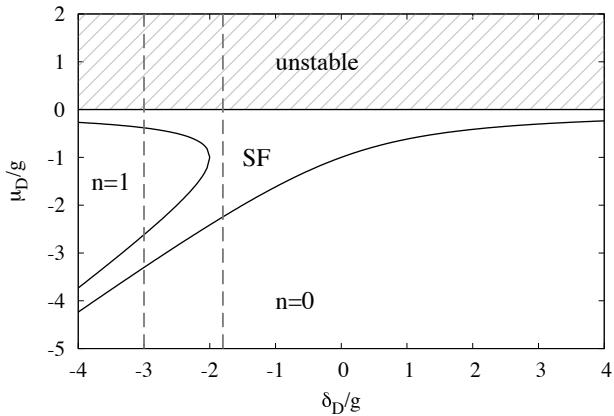} 
\caption{\small Quantum phase diagram of the JCHM at infinite bandwidth for fixed $\mu_D,\delta_D$ showing the
transition from the vacuum ($n=0$) to a superfluid state as well as the existence of a single Mott lobe with $n=1$. 
The vertical dashed lines indicate the values used in Fig.~\ref{fig:critical-T}.}
\label{dicke-t0}
\end{figure}

The zero temperature phase diagram is shown in Fig.~\ref{dicke-t0}.
As $\mu_D \to 0^-$, the critical coupling strength goes to zero and so
a superradiant region is always seen near $\mu_D=0$.  For $\delta_D <
0$, more complicated behaviour vs $\mu_D$ occurs --- two
superradiant (superfluid) regions exist, separated by a normal state.
Near $\mu_D =\delta_D$ there is a high susceptibility of the two-level
systems towards polarisation and hence a superradiant region exists
near $\mu_D =\delta_D$.  The normal state for $\mu_D > \delta_D$ has
all two-level systems inverted and thus corresponds to a Mott lobe
with one excitation per cavity.

With this identification of the Mott lobe, the tip of the lobe
can be found as the value of $\delta_D$ for which two branches of the
phase boundary merge.  Since the $n=1$ lobe occurs within the region
$\delta_D - \mu_D < 0$ the modulus sign in the expression for the
critical $g$ can be removed and the phase boundary of this lobe
becomes
\begin{equation}
  \label{eq:2}
  \mu_D = \frac{1}{2} \left( \delta_D \pm \sqrt{\delta_D^2 - 4g ^2} \right)
\end{equation}
hence the lobe tip is at $\delta_D = - 2 g, \mu_D=-g$ as is clear from
Fig.~\ref{dicke-t0}. As we are going to show in the next section, this result
exactly agrees with those obtained from the slave boson theory derived 
in the previous section.

\subsection{Excitation spectra}
\label{sec:excitation-spectra}

We now look at the excitation spectra.  These can be found by writing an
effective action for the Dicke model and expanding around the saddle
point corresponding to the mean-field solution~\cite{Eastham2001,KE05}.
Identifying the poles of the Green's function then gives the
excitation spectrum.  In general, they can be written
as
\begin{eqnarray}
  \label{eq:4}
\epsilon_{\pm}({\bf k})=\sqrt{A({\bf k})\pm\sqrt{A({\bf k})^2- B({\bf k})}}
\end{eqnarray}
with
\begin{eqnarray}
  \label{eq:5}
A({\bf k})&=&(E^2+\tilde{\omega}_{\bf k}^2 + 2\tilde{\omega}_x\tilde{\omega}_{{\bf k}=0})/2,\\
\label{eq:6}
B({\bf k})&=&(\tilde{\omega}_{\bf k}-\tilde{\omega_0})(E^2\tilde{\omega}_{\bf k}-\tilde{\omega}_x^2\tilde{\omega}_{{\bf k}=0}).
\end{eqnarray}
In the normal phase with $\psi_0=0$, these expressions simplify to
\begin{eqnarray}
  \label{eq:3}
\epsilon_{\pm}({\bf k})=\frac{1}{2}\left(J{\bf k}^2+\delta_D-2\mu_D\pm\sqrt{Q({\bf k})}\right)
\end{eqnarray}
with
\begin{eqnarray}
  Q({\bf k})=\left(J{\bf k}^2-\delta_D\right)^2 + 4g^2 \mathrm{sgn}(\delta_D-\mu_D)\,,
\end{eqnarray}
where $\mathrm{sgn}(x) = x/|x|$.  

It follows that the gap of the amplitude mode is given by
$\Delta_a=|\delta_D - 2 \mu_D| = \sqrt{\delta_D^2-4g^2
  \mathrm{sgn}(\mu_D - \delta_D)}$.  Thus, this gap is non-zero
everywhere on the $n=1$ to superradiant boundary (where $\mu_D <
\delta_D$). At the tip of the $n=1$ to superradiant boundary, i.e., at
$\delta_D = -2 g$, there is a vanishing gap of the amplitude mode.
Similarly there is vanishing sound velocity of the gapless mode
everywhere on the phase boundary, except at the tip of the lobe where
a vanishing gap leads to a linear dispersion with sound velocity
$c_s=\sqrt{Jg}$.

\section{Connection between the two limits}
\label{sec:connection}

\subsection{Lobe tips in quantum theory}
\label{sec:lobe-tips-quantum}
In order to compare the two approaches valid in the weak and strong
correlation limits, we evaluate the slave boson theory at infinite
bandwidth $J\rightarrow \infty$ while keeping $\mu_D$ and $\delta_D$ fixed,
corresponding to infinite negative detuning.  With the expressions in
the appendix we obtain for the phase boundary $\mu_D=(\delta_D\pm\sqrt{\delta_D^2-4 g^2})/2$
matching exactly Eq.~(\ref{eq:2}), with the tip of the lobe at
$\delta_D=-2g$ and $\mu_D=-g$.  Here, one mode remains gapless, the
other maintains a gap $\Delta_a = \sqrt{\delta_D^2-4g^2}$ away from the tip.
The sound velocity vanishes everywhere except at the tip of the lobe where
$c_s=\sqrt{Jg-g^2/(2D)}\approx \sqrt{Jg}$. These results agree exactly with those obtained from a weak-coupling mean-field theory 
in the previous section.

Thus, the weak-coupling mean-field theory describes correctly the superfluid-Mott insulator
transition of the lowest Mott lobe with $n=1$ in the limit of large hopping and large
negative detuning. In Fig.~\ref{fig1} one can see the reason for this success:
the size of the lowest Mott lobe increases for large $J$ and large negative $\delta$, while the size of all other lobes 
decreases. Thus, only one lobe survives in Fig.~\ref{dicke-t0}. All other modes are pushed towards $\mu_D=0$
and vanish.
The success of the weak-coupling theory for the JCHM is in strong contrast to the Bose-Hubbard model, where
a Bogoliubov-like theory describing weakly interacting atomic BEC`s,
fails to predict the existence of Mott lobes and gapped Higgs modes at
weak interactions \cite{Oosten2001}. 
In the polariton picture, one source of this
difference is clear: the mean field theory of the Dicke model has
normal modes that result from hybridisation of the photon modes and
spin-waves of the two-level systems, leading to two polariton
branches~\cite{pekar58,hopfield58}.  
The nature of these branches depends on the detuning parameter $\delta$, which does not exist for the BHM.
At the phase boundary, one of these branches turns into the Higgs mode, the other corresponds to the Goldstone mode
in Fig.~\ref{fig2}. It is interesting to note that the agreement of these results occurs even despite the upper polariton state
having been eliminated in the slave-boson theory. To understand how this can be so, one should note that this elimination
is in terms of the modes defined on a single cavity, whereas the weak coupling theory describes polaritons arising from the 
delocalised photon mode, thus corresponding to different mixtures of qubit and photon as well as phase and amplitude
degrees of freedom.

\subsection{Finite temperature phase transition and mean-field theory}
\label{sec:finite-temp-phase}
\begin{figure}[t]
  \centering
  \includegraphics[width=0.45\textwidth]{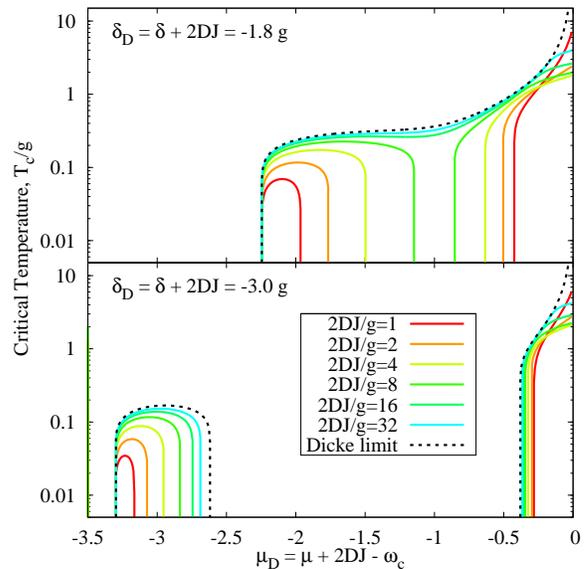}
  \caption{Critical temperature calculated from mean-field theory of
    the JCHM (solid lines) corresponding to a generalization of
      Eq.~(\ref{boundary}) to finite temperature, see
      e.g.~\cite{Nietner2012} and from the Dicke model (dashed
    lines).  In the large bandwidth limit, these match.  Top panel is
    plotted for a detuning $\delta_D$ which misses the n=1 Mott lobe
    in the Dicke limit, and has a single superfluid phase above a
    critical $\mu_D$.  Lower panel has a more negative detuning
    $\delta_D$ so the Mott lobe exists in the Dicke limit. Note, that the dashed lines
    at $T=0$ correspond to the vertical cuts through the phase diagram in Fig.~\ref{dicke-t0}.}
  \label{fig:critical-T}
\end{figure}

As noted above, the mean field theory of the Dicke model is only
correct in the limit of large bandwidth (small photon mass), where the
density of states for fluctuation corrections vanishes.  In the
context of the Dicke model with a quadratic dispersion, a heavier mass
does not mean a finite bandwidth, but does mean fluctuation
corrections are important.  In the absence of a finite bandwidth,
tight-binding (strong correlations) approaches cannot so easily be
applied.  An alternate approach to account for the fluctuation
corrections is to consider the fluctuation correction to the effective
action, following Nozi\`eres \&
Schmitt-Rink~\cite{Nozieres1985,randeria}.  When such an approach is
carried out for the Dicke model~\cite{KE05} one notable effect is that
even for $\Delta > -2 g$, there can be multiple disconnected normal
regions, i.e., fluctuations can suppress the superradiant phase,
leading to the appearance of the $n=1$ Mott lobe.  This suppression of
superradiance is most clearly seen by considering the critical
temperature $T_c$ for the superradiance transition.  Including
fluctuations, this critical temperature can be suppressed to $T_c=0$,
leading to two disconnected superradiant ``bubbles'' in Fig.\ref{fig:critical-T}.

The suppression of $T_c$ due to fluctuations has a natural
interpretation in the context of the JCHM.  Including fluctuations
corresponds to increasing the photon mass, or decreasing the effective
hopping.  For the JCHM, it is self-evident that reducing hopping will
lead to the suppression of superfluidity.  
This is shown in Fig.~\ref{fig:critical-T} which compares the critical temperature
coming from the standard decoupling mean field theory (as opposed to weak-coupling mean-field theory)
of the JCHM to the critical
temperature of the corresponding Dicke model. These clearly show how
the Dicke limit is recovered (both panels).
Note, that the quantum phase diagram as obtained from decoupling mean-field theory
agrees exactly with the results of the slave-boson theory.
\section{Conclusions and Outlook}
\label{sec:conclusion}
In this paper, we have shown that a remarkable relation exists between
a slave-boson theory for the JCHM and a weak-coupling mean-field
theory for the multi-mode Dicke model. We found that both theories
match exactly in the limit of infinite bandwidth and negative
detuning. In this special limit a single Mott lobe survives and thus a
weak-coupling mean-field theory is capable of describing the
superfluid-Mott insulator transition with correct critical exponents.
So far, our work was based on the equilibrium assumption, i.e., a
chemical potential was introduced in order to fix the number of
polaritons inside the cavity-array.  It would be very
interesting to see whether predictions of the weak-coupling mean-field
theory for the driven dissipative generalizations of the Dicke model
\cite{Szymanska2006,Keeling2013,Dimer2007,Baumann2010,Nagy2010,Keeling2010,Nagy2011c,Konya2012,Bhaseen2012,Oztop2012a,Torre2012}
also allow predictions for the fate of the SF-MI transition under
non equilibrium conditions.
\begin{acknowledgements}
This work was supported by a SNF Ambizione award 
(Swiss National Science Foundation).
\end{acknowledgements}
\appendix
\setcounter{section}{1}
\section*{Appendix: Abbreviation in the slave-boson formalism }
The matrix elements of the two-by-two matrices in Eq.~(\ref{eq:sb-heff}) are given by
\begin{eqnarray}
g_{11}&=&(\cos(2\theta)U_-+Jz\sin(2\theta)^2C_+^2)/2\nonumber\\
&&- J(k_0^2+k_1^2)\epsilon_q/2\\
g_{22}&=&(U_+-\sin(\theta)^2 U_-+Jz\sin(2\theta)^2C_+^2/2)/2\nonumber\\
&& - J(h_0^2+h_1^2)\epsilon_q/2\\
g_{12}&=&-\cos(\theta)(\sin(2\chi)(\epsilon_1-\epsilon_{-1}) + 8Jz\sin(\theta)^2C_+C_-)/4\nonumber\\
&&+J(h_1 k_0+h_0 k_1)\epsilon_q/2\\
g_{21}&=&g_{12}
\end{eqnarray}
and
\begin{eqnarray}
f_{11}&=&J k_0 k_1 \epsilon_q\\
f_{22}&=&J h_0 h_1 \epsilon_q\\
f_{12}&=&-J (h_0 k_0+h_1 k_1) \epsilon_q/2\\
f_{21}&=&f_{12}
\end{eqnarray}
with the definitions
\begin{eqnarray}
U_-(\chi)&=&\epsilon_{-1}\sin(\chi)^2+\epsilon_1\cos(\chi)^2-\epsilon_0\\
U_+(\chi)&=&\epsilon_{-1}\cos(\chi)^2+\epsilon_1\sin(\chi)^2-\epsilon_0\\
C_-(\chi)&=&f_1\sin(\chi)-f_0\cos(\chi)\\
C_+(\chi)&=&f_1\cos(\chi)+f_0\sin(\chi)
\end{eqnarray}
and
\begin{eqnarray}
h_0&=&f_0\cos(\theta)\cos(\chi)\\
h_1&=&f_1\cos(\theta)\sin(\chi)\\
k_0&=&f_1\cos(\theta)^2\cos(\chi)-f_0\sin(\theta)^2\sin(\chi)\\
k_1&=&f_1\sin(\theta)^2\cos(\chi)-f_0\cos(\theta)^2\sin(\chi)
\end{eqnarray}
The fluctuation correction to the ground-state energy in Eq.~(\ref{fluct}) reads
\begin{eqnarray}
\mathcal{E}(\theta,\chi)/N_s&=&-(3/4) J z \sin(2\theta)^2 C_+(\chi)^2 - U_+(\chi)/2\nonumber\\
&&- (\cos(2\theta)-2\sin(\theta)^2) U_-(\chi)/2      
\end{eqnarray}
Finally, the expressions describing the excitation spectra in Eq.~(\ref{spectterms}) are given by
\begin{eqnarray}
A({\bf k})&=&2(g_{11}^2+g_{22}^2+2 g_{12}\, g_{21}- f_{11}^2 - f_{22}^2 - 2 f_{12}\, f_{21})\nonumber\\
\\
B({\bf k})&=&16\left[(g_{11} - f_{11})(g_{22} - f_{22}) - (g_{12} - f_{12})^2\right]\nonumber\\
&&\times \left[(g_{11} + f_{11})(g_{22} + f_{22})-(g_{12} + f_{12})^2\right]\nonumber\\
\end{eqnarray}
\\

%

\end{document}